# From High-Entropy Ceramics to Compositionally-Complex Ceramics: A Case Study of Fluorite Oxides

Andrew J. Wright[a], Qingyang Wang[b], Chuying Huang[a], Andy Nieto[c], Renkun Chen[b], Jian Luo[a,*]

[a]Department of NanoEngineering, University of California, San Diego, La Jolla, California, USA

[b]Department of Mechanical & Aerospace Engineering, University of California, San Diego, La Jolla, California, USA

[c]Department of Mechanical and Aerospace Engineering, Naval Postgraduate School, Monterey, California, USA

**Abstract**

Using fluorite oxides as an example, this study broadens high-entropy ceramics (HECs) to compositionally-complex ceramics (CCCs) or multi-principal cation ceramics (MPCCs) to include medium-entropy and/or non-equimolar compositions. Nine compositions of compositionally-complex fluorite oxides (CCFOs) with the general formula of $(Hf_{1/3}Zr_{1/3}Ce_{1/3})_{1-x}(Y_{1/2}X_{1/2})_xO_{2-\delta}$ ($X$ = Yb, Ca, and Gd; $x$ = 0.4, 0.148, and 0.058) are fabricated. The phase stability, mechanical properties, and thermal conductivities are measured. Compared with yttria-stabilized zirconia, these CCFOs exhibit increased cubic phase stability and reduced thermal conductivity, while retaining high Young's modulus (~210 GPa) and nanohardness (~18 GPa). Moreover, the temperature-dependent thermal conductivity in the non-equimolar CCFOs shows an amorphous-like behavior. In comparison with their equimolar high-entropy counterparts, the medium-entropy non-equimolar CCFOs exhibit even lower thermal conductivity ($k$) while maintaining high modulus ($E$), thereby achieving higher $E/k$ ratios. These results suggest a new direction to achieve thermally-insulative yet stiff CCCs (MPCCs) via exploring non-equimolar and/or medium-entropy compositions.

*Correspondence should be addressed to J.L. (email: jluo@alum.mit.edu)

**Summary of Novel Conclusions:**

This study broadens “high-entropy ceramics” to “compositionally-complex ceramics”. Using fluorite oxides as an example, we show that lower thermal conductivity and higher stiffness-to-conductivity ratios are achieved in medium-entropy non-equimolar compositions.

**Highlights:**

- Nine compositionally-complex fluorite oxides (CCFOs) are made and investigated.
- CCFOs exhibit reduced thermal conductivity and increased cubic phase stability.
- Lower thermal conductivity is achieved in medium-entropy non-equimolar CCFOs.
- High modulus and hardness retain in CCFOs with reduced thermal conductivity.
- Non-equimolar CCFOs exhibit amorphous-like *T*-dependent thermal conductivity.

## 1 Introduction

High-entropy alloys (HEAs) have been developed since 2004 with the seminal works of Yeh et al. [1] and Cantor et al. [2]. A key advantage of HEAs is a vast compositional space to enable improved and tailorable properties. The initial focus is on equiatomic compositions, where the configurational entropy is maximized as $\Delta S_{conf} = RlnN$ per mole, where $N$ (typically ≥ 5) is the number of elements and $R$ is gas constant. In the physical metallurgy community, HEAs have been expanded to multi-principal element alloys (MPEAs) and complex-concentrated or compositionally-complex alloys (CCAs) to include medium-entropy and non-equiatomic compositions, which display excellent strength and ductility that can outperform their equiatomic counterparts [3–5].

Since 2015, high-entropy ceramics (HECs) have been made in bulk form, including various high-entropy oxides [6–10], borides [11,12], carbides [13–15], nitrides [16], and silicides [17,18]. Unlike metallic HEAs of simple FCC, BCC or HCP structures, HECs typically only have high-entropy mixing at one (or multiple) cation sublattice(s). Interestingly, single-phase high-entropy aluminides (intermetallic compounds with high-entropy mixing in only one of the two metal sublattices, albeit substantial anti-site defects and disorder in the other Al-rich sublattice) have also been made [19], thereby bridging the gap between HEAs and HECs. The HECs have been demonstrated on multiple occasions to exhibit lower thermal conductivities than their constituents because of stronger phonon scattering [9,20,21]. In some sense, HECs exhibit amorphous-like thermal conductivity or "phonon glass" behavior.

Yttria-stabilized zirconia (YSZ) with different yttria contents has been widely used [22,23] because of its superior properties, such as ionic conductivity [24] and fracture toughness [23]. Particularly, YSZ has been used as thermal barrier coatings (TBCs) for gas-turbine engines due to low thermal conductivity [25]. In general, there are great interests to search for new materials with low thermal conductivity and high stiffness for potential applications as thermally-insulating protective coatings. In this regard, it is natural to explore "high-entropy" versions of YSZ, which can offer further reduced thermal conductivities. High-entropy fluorite oxides (HEFOs) have been made on three occasions. Djenadic et al. synthesized single-phase high-entropy rare earth oxide powders such as $(Ce_{0.2}La_{0.2}Pr_{0.2}Sm_{0.2}Y_{0.2})O_2$ [7]. Gild et al. first reported the fabrication of a series of eight densified bulk YSZ-like HEFOs, e.g., $(Zr_{0.2}Ce_{0.2}Hf_{0.2})(Y_{0.2}Gd_{0.2})O_2$, with equimolar $HfO_2$-$ZrO_2$-$CeO_2$ as the matrix plus one or two stabilizers (selected from $Y_2O_3$, $Yb_2O_3$, $Gd_2O_3$, and CaO) [9]. Chen et al. also reported a high-entropy $(Zr_{0.2}Hf_{0.2}Ce_{0.2}Sn_{0.2}Ti_{0.2})O_2$ (albeit some compositional inhomogeneity indicated by EDS mapping, particularly for Ce) with low thermal conductivity [26]. Moreover, Gild et al. showed 25 – 50 % reductions in the thermal conductivity of the YSZ-like HEFOs in comparison with YSZ. To date, all studies of HECs (including all HEFOs) only explored equimolar or near equimolar compositions.

In this study, we further expand the work to non-equimolar compositions to achieve better performance such as lower thermal conductivity for TBCs, which represent a new direction to further broaden the compositional space. A somewhat surprising (but fully explainable) and scientifically interesting result is that the thermal conductivity can be reduced further by exploring non-equimolar compositions.

## 2 From High-Entropy Ceramics (HECs) to Compositionally-Complex Ceramics (CCCs)

HEAs were initially defined as alloys containing five or more elements in equimolar ratios by Jien-Wei Yeh in 2004 [1]**.** This definition has been broadened to MPEAs or CCAs to include alloys containing multiple principal elements of 5 at. % to 35 at. % [27,28]. Alternatively, Yeh et al. propose to classify alloys based on their ideal molar mixing entropies as low-entropy ($\Delta S_{conf} < 0.69R$), medium-entropy ($0.69R \leq \Delta S_{conf} < 1.61R$), and high-entropy ($\Delta S_{conf} \geq 1.61R$) alloys [29]. Others proposed medium-entropy alloys to have $R < \Delta S_{conf} < 1.5R$ (so that $\Delta S_{conf} \geq 1.5R$ per mole for HEAs) [30]. The latter seems to be a more realistic classification of medium- and high-entropy compositions that we adopt. Note that here $\Delta S_{conf}$ is calculated assuming ideal random mixing, albeit clustering and local ordering in the real solid solutions can reduce the actual mixing entropy.

Moreover, the concept of entropy-stabilized phases was proposed [31]**.** A system is said to be entropy stabilized if the mixing entropy alone can overcome the positive enthalpic barrier to stabilize the solid-solution phase. In 2015, Rost et al. reported an entropy-stabilized oxide, $(Mg_{1/5}Ni_{1/5}Co_{1/5}Cu_{1/5}Zn_{1/5})O$, of the rocksalt structure [6]**.** The entropy-stabilized ceramics are generally a subset of HECs, but entropy stabilization can also occur in cases that are not officially "high entropy" based on the above definitions, as schematically shown in Figure 1.

It is worth noting that (medium-entropy) multi-cation ceramics have already been studied by the ceramic community, e.g., $BaZr_{0.1}Ce_{0.7}Y_{0.2-x}Yb_xO_{3-\delta}$ for solid oxide fuel cell applications [32], albeit that they are typically not near equimolar compositions. However, efforts to develop HECs have so far been focused on equimolar or near equimolar compositions. Similar to the expansion from HEAs to MEPAs and CCAs in the physical metallurgy community, there is a potential benefit to broaden the realm of HECs to explore non-equimolar and/or medium-entropy compositions, which represent an even larger compositional space.

Following the definitions of metallic HEAs, HECs typically refer to equimolar or near-equimolar compositions of five or more principal cations, which usually produce >1.5 $R$ per mol. ideal configuration entropy $\Delta S_{conf}$, on at least one sublattice. However, like in the metallic HEAs, some researchers [13] also consider equimolar, four-cation compositions (with 1.39 $R$ per mol. $\Delta S_{conf}$) as HECs.

Here, we propose to loosely define a class of medium- to high-entropy compositionally-complex ceramics (CCCs or $C^3$), which can alternatively and equivalently be named as "multi-principal cation ceramics (MPCCs)," as single-phase ceramic solid solutions with at least three principal (meaning typically 5%-35% on a sublattice) cations (mixing on at least one sublattice, if there are more than one cation sublattice), which should generally have $> R$ per mole of metal cations in the ideal $\Delta S_{conf}$ (on at least one sublattice).

Thus, HECs and entropy-stabilized ceramics [33,34] are subsets of CCCs or MPCCs, which also include three- or four-component medium-entropy ceramics and/or non-equimolar compositions (Figure 1). It is important to note that the definitions of high- and medium-entropy compositions are most subjective; thus, the respective boundaries are only loosely defined (so that different approaches based on the configuration entropy or the number and percentages of principal cations will lead to somewhat different cutoffs of the classifications). We should note that most non-equimolar and/or medium-entropy ceramics (based on the above definitions) also represent a largely unexplored compositional space.

Pursuing along the line to broaden HECs to CCCs or MPCCs to include medium-entropy non-equimolar compositions using fluorite oxides as an exemplar, this study explores the fabrication, phase stability, mechanical properties, and thermal conductivities of nine compositions of compositionally-complex fluorite oxides (CCFOs). These CCFOs have the general formula of $(Hf_{1/3}Zr_{1/3}Ce_{1/3})_{1-x}(Y_{1/2}X_{1/2})_xO_{2-\delta}$ where $X$ = Yb, Ca, and Gd, respectively, and $x$ = 0.4, 0.148, and 0.058, respectively. Notably, the medium-entropy non-equimolar CCFOs exhibit even lower thermal conductivity ($k$) and higher modulus-to thermal conductivity ($E/k$) ratios than their equimolar high-entropy counterparts. Thus, this study suggests a new direction to design, fabricate, and test stiff and insulative CCFOs (or explore CCCs or MPCCs in general) via exploring non-equimolar and/or medium-entropy compositions that also represent a largely unexplored, vast, compositional space (in addition to HECs).

## 3 Experimental Procedures

### 3.1 Materials and Synthesis

Three variants of three different cation combinations (i.e., nine compositions) of CCFOs were explored in this study. The "H" variant represents high cubic stabilizer concentration and equimolar composition of the maximum configurational entropy, i.e., $(Hf_{1/5}Zr_{1/5}Ce_{1/5}Y_{1/5}X_{1/5})_xO_{2-\delta}$, where $X$ = Yb for Cation Combination #4, Ca for Cation Combination #5, and Gd for Cation Combination #7 (based on the numerical series used in the prior report by Gild et al. [9]). The "M" and "L" variants represent medium and low stabilizer concentrations, and their total concentration of the two stabilizers (Y and $X$) matches those in 8 mol. % YSZ and 3 mol. % YSZ, respectively, with equimolar Hf, Zr, and Ce: i.e., $(Hf_{1/3}Zr_{1/3}Ce_{1/3})_{1-x}(Y_{1/2}X_{1/2})_xO_{2-\delta}$, where $x$ = 0.148 for the "M" variant and $x$ = 0.058 for the "L" variant, respectively. Note that, e.g., 8YSZ represents $(Y_2O_3)_{0.08}$ $(ZrO_2)_{0.92}$, which is equivalent to $Y_{0.148}Zr_{0.852}O_{2-\delta}$. The nine CCFO compositions, along with 3YSZ and 8YSZ, are shown in Table 1.

Each CCFO was synthesized from commercially-purchased $HfO_2$ (99.99 %, 61 – 80 nm, US Research Nanomaterials, Inc.), $ZrO_2$ (99 %, 15 – 25 nm, Alfa Aesar), $CeO_2$ (99.5 %, 15 – 30 nm, Alfa Aesar), $Y_2O_3$ (99.995 %, 25 – 50 nm, Alfa Aesar), $Yb_2O_3$ (99.9 %, < 50 nm, MTI Corporation), CaO (98 %, ≤ 160 nm, Sigma-Aldrich), and $Gd_2O_3$ (99.99 %, 20 – 40 nm, Alfa Aesar) nanopowders. The powders were weighed based on the targeted stoichiometries and placed inside a YSZ planetary mill jar with ethanol. YSZ grinding media was also added to the jar with the ball-to-powder mass ratio of approximately 10:1. The powders were then wet-milled in a planetary mill (Across International PQ-N04, NJ, USA) operating at 300 RPM for 24 hours. The milled slurry was transported to an 85°C drying oven to dry overnight. After drying, the powders were lightly ground using a mortar and pestle to break up any agglomerates that formed during the drying process.

The powders were consolidated into pellets by using a 10 mm diameter graphite die inside a spark plasma sintering (SPS) machine (Thermal Technologies, CA, USA) to sinter at 1650°C for 5 min under 50 MPa, with a heating ramp rate of 100°C/min. The powder was loaded into a 10 mm graphite die that was internally lined with 125 μm thick graphite foil. The die was then insulated by covering all surfaces with graphite felt. The chamber was pumped down to at least 20 mTorr. The chamber was under this vacuum until approximately 700°C, when argon was filled into the chamber and remain for the rest of the sintering process. After the five-minute dwell at 1650°C, the pressure was immediately released, and power supply was turned off. Following cooling, the pellets were decarburized and homogenized at 1500°C for 24 hours and allowed to naturally cool in the furnace. Note that the 1500°C, 24 hours annealing in air is essential to oxidize the SPS specimens made in a reduced environment and homogenize the compositions to form single-high entropy oxide phases. The same fabrication procedure was used to sinter 3YSZ and 8YSZ as our benchmark specimens.

## 3.2 Characterization

### 3.2.1 Scanning Electron Microscopy (SEM)

A scanning electron microscope (SEM, FEI Apreo, OR, USA) equipped with an energy dispersive X-ray spectroscopy (EDS) detector was used to investigate the microstructure and homogeneity of polished samples. Before SEM examination, the samples were mounted and polished down to 40 nm with colloidal silica suspension and then coated with a thin layer of iridium to minimize charging.

### 3.2.2 X-ray Diffraction (XRD), Density, and Phase Stability

Following the homogenization step, each sample was pulverized into a powder with a mortar and pestle. The crystal structure of the sample was examined through an XRD (Rigaku Miniflex II, Tokyo, Japan) operating at 30 kV and 15 mA with a scan step of 0.01° 2θ and a dwell time of 3 s. The XRD was calibrated (peak position and breadth) with $LaB_6$ purchased from NIST (SRM 660b). XRD analysis was conducted through MDI Jade 6. It is assumed that the material exhibits the stoichiometric amount of oxygen vacancies and that Ce is in 4+ oxidation state. The microstrain of each material was calculated by Williamson-Hall analysis [35].

The bulk density was calculated by both the ASTM C373-18 Standard boiling method [36] and digital image threshold processing (ImageJ) of SEM micrographs, which were in close agreements.

Phase stability was tested by long furnace annealing (for 24 h) followed by rapid air quenching. After air quenching to room temperature, the pellets were ground into a powder and examined by XRD.

### 3.2.3 Raman Spectroscopy

A confocal Raman microscope (Renishaw inVia, England, UK) equipped with a 40 mW, 488 nm $Ar^+$ laser, and a 1 μm spot size was used to characterize the tetragonality of the selected specimens. The power was kept below 2 mW to avoid local heating. The experiments were performed on compacted powder samples after decarburization. The data were collected for 10 seconds and averaged over 10 scans.

### 3.2.4 Mechanical Properties

The Young's modulus was determined through a pulse-echo sonic resonance setup by an oscilloscope (TDS 420A, Tektronix, USA) following the ASTM C1198-09 Standard [37]. The material's longitudinal ($u_L$) and transverse ($u_T$) wave speeds were determined from the measurements and then were used to calculate Poisson's ratio ($\nu$) and Young's Modulus (here we used "$E_{measured}$" to represented the directly measured modulus with minor porosity) according to the following Eqs. (1) and (2) below:

$$\nu = \frac{u_L^2 - 2u_T^2}{2(u_L^2 - u_T^2)} \tag{1}$$

$$E_{measured} = 2u_T^2\rho(1 + \nu) \tag{2}$$

Subsequently, Young's modulus was corrected for porosity by:

$$E_{measured} = E(1 - 2.9P), \tag{3}$$

where $E$ (i.e., $E_0$ in the equations in Ref. [38]) is the corrected Young's modulus for a fully-dense specimen [38] that we report in Table 1 and used for all discussion.

The mechanical property of the CCFOs was also probed by nanoindentation by using a Hysitron TI 950 nanoindenter (MA, USA) based on the widely accepted Oliver-Pharr method [39,40]. A 150 nm diamond Berkovich tip was used. The max loading force for indentation was 9 mN. The loading profile consisting of 10 s ramp to the maximum load, followed by a 5 s holding period, and lastly a 10 s unload. All measurements were performed at room temperature on polished surfaces. A 5 × 5 grid with 5 μm spacing between each point was used for the indent locations. This grid was selected for three nearby areas on the sample for a total of 75 indents. Obvious outliers (e.g., an indentation on pores or scratches) were removed from the analysis after inspection.

### 3.2.5 Thermal Conductivity

Thermal conductivity measured from room temperature to 1000°C was determined by the product of thermal diffusivity ($\alpha$), density ($\rho$), and specific heat capacity ($c_p$) based on the following equation:

$$k_{measured} = \alpha \rho c_p \tag{4}$$

The thermal diffusivity was measured by laser flash analysis (LFA 467 *HT HyperFlash*, NETZSCH, Germany). The heat capacity was estimated through Neumann-Kopp rule by the rule of mixtures of the constituent oxides [41]. Finally, the conductivity was corrected for porosity based on the following equation:

$$k_{measured} = k(1-P)^{3/2} \tag{5}$$

Here, $k$ (i.e. $k_0$ in equations in Ref. [42]) is the corrected thermal conductivity for a fully-dense specimen [42]. Again, here we adopt $k_{measured}$ to represent the measured thermal conductivity of specimens with minor porosity and reserve $k$ for the corrected value reported in Table 1 and discussed in all places.

## 4 Results and Discussion

### 4.1 XRD

XRD was performed on all 11 compositions and the results are shown in Figure 2. The CCFOs are compared to 8YSZ and 3YSZ that were fabricated by the same procedure.

8YSZ shows the expected cubic fluorite structure. 3YSZ shows a mixture of the tetragonal, cubic, and monoclinic phases. After annealing at 1500°C for 24 hours, 3YSZ exhibits the metastable tetragonal phase; however, it is known that this phase can transform into tetragonal and cubic phases after prolonged high-temperature annealing [43]. The tetragonal phase could then transform into the monoclinic phase during cooling. Due to the presence of multiple phases, the properties of 3YSZ were not measured (because we are only interested in the intrinsic properties of the single-phase material).

For the nine CCFOs annealed at 1500°C for 24 hours (furnace-cooled), all the "H" and "M" compositions, except for 5H, possess a single cubic fluorite phase ($Fm\bar{3}m$, space group #225). Composition 5H still predominantly contains a fluorite phase; however, there is a secondary $Ca(Hf/Zr)O_3$ perovskite phase.

Among the "L" variants annealed at 1500°C for 24 hours, 4L and 7L show a single body-centered tetragonal (BCT) phase, similar to the tetragonal YSZ (P42/nmc, space group #137). Composition 5L (with Y and Ca as the stabilizers) exhibits a single cubic fluorite phase. The addition of every $Ca^{2+}$ cation generates an $O^{2-}$ oxygen vacancy, doubling that produced by adding a 3+ cation. Since it is known that the formation of the cubic fluorite structure strongly depends on the concentration of oxygen vacancies [44], the additional oxygen vacancies in 5L with $Ca^{2+}$ stabilizers likely stabilize this cubic phase.

On examination of the XRD spectra, the peaks widen as the stabilizer concentration decreases. In general, peak broadening can arise from small crystallite/domain size, residual strains, and inhomogeneity. The residual strains and small crystallite size effects can be excluded in this case based on the Williamson-Hall analysis and long annealing times. Furthermore, our samples are compositionally homogeneous (as shown by the EDS analysis later). Thus, we propose that the peak broadening observed here is likely resulted from partial/initial phase transformation. Specifically, an FCC to BCT transition will result in peak splitting when the *c*/*a* ratio in the BCT cell deviates from the $\sqrt{2}$ of the cubic phase. However, due to XRD instrumental broadening, the initial small splitting looks similar to peak broadening.

Here, the microstrain parameter ($\varepsilon$) extracted from the Williamson-Hall plot can be used as a proxy for the degree of broadening due to the presence of the tetragonal phase, in which the effects can be seen in variant "L". As the stabilizer concentration increases, the broadening parameter decreases, suggesting that the tetragonality of the sample decreases.

Table 1 summarizes the relevant data obtained from XRD. Compositions 5H and 3YSZ were not single-phase (so their properties were not further measured). Compositions 4L and 7L were classified as BCT. The $c/a$ ratio for 4L and 7L are 1.423 and 1.421, respectively. When $c/a = \sqrt{2} \approx 1.414$, the phase is equivalent to FCC (fluorite). Thus, 4L and 7L only exhibit moderate tetragonality. The rest of the CCFOs exhibit the cubic fluorite structure according to XRD (albeit some trace tetragonality that results in peak broadening). The lattice parameter for each composition increases with the increasing stabilizer concentration due to the larger sizes of the stabilizer atoms compared to the host atoms. Assuming XRD peak broadening is due to the initial FCC to BCT phase transition, $\varepsilon$ increases as the stabilizer concentration decreases. The tetragonality at lower stabilizer concentrations is expected, similar to that observed in YSZ [43].

### 4.2 Raman Spectroscopy

Raman spectroscopy was used to investigate the presence of moderate tetragonality, which is more sensitive than XRD. This technique is susceptible to local short-range ordering. Figure 3 shows the Raman spectra for 8YSZ, 4L, 4M, and 4H. A pure cubic specimen should exhibit a single sharp peak around 465 $cm^{-1}$ [45–47]. In the specimens measured, there are six peaks which correspond to the tetragonal symmetry [48–50]. Although nearly all the XRD spectra show a strong cubic structure, the short-range order suggests some tetragonality in all specimens, which decreases as the stabilizer concentration increases. Some researchers have suggested that this tetragonality may arise from distorted C-type/pyrochlore micro/nanodomains within the system [51–53]. The broadened peaks are likely due to both the decreasing tetragonality and a highly disordered structure. It is important to note that in this case (excluding the composition series 5), variants "H" and "M" are cubic, and "L" is tetragonal according to XRD. Although Raman spectra indicate some tetragonality in all nominally cubic phase, we used XRD as our primary criterion to differentiate cubic and tetragonal phases (as the tetragonality revealed by Raman is small, yet universal).

In summary, XRD suggested that 4L is tetragonal (in long-range order), while both 4M and 4H are cubic (exhibits long-range cubic order) according to XRD, but Raman spectroscopy revealed some tetragonality or possible presence of some short-range tetragonal order. In fact, 8YSZ is commonly known as a cubic phase. However, it is also known, and verified here, that the Raman spectra can reveal some tetragonality (short-range tetragonal order) even in the well-known cubic 8YSZ [48–51].

### 4.3 Phase Stability

The stability of 4L, 4M, and 4H (or $(Hf_{1/3}Zr_{1/3}Ce_{1/3})_{1-x}(Y_{1/2}Yb_{1/2})_xO_{2-\delta}$) was examined further by utilizing specimens quenched from different temperatures. Here, each sample was first annealed at 1500°C for approximately 24 h in air (that is essential to oxidize and homogenize the specimens made by SPS in a reduced environment), and then rapidly air quenched to preserve the crystal structure. XRD revealed that all variants exhibited sharp fluorite peaks, including 4L that exhibited tetragonal phase in the slow furnace cooled specimen (Figure 1 for slow furnace cooling vs. Figure 4 for air-quenched specimen). This indicates that 4L underwent a cubic to tetragonal phase change during furnace cooling.

After quenching at 1500°C to form a single cubic phase, each sample was then annealed again at 1200°C – 1400°C for 24 h – 48 h. It was found that variants "H" and "M" remained cubic with sharp peaks for all temperatures examined and are shown in Figure S1 in Suppl. Data. Variant "L", however, underwent a BCT to FCC phase transition around 1400°C – 1500°C. The XRD spectra for quenched 4L from all temperatures are shown in Figure 4(a). The expanded spectra in Figure 4(b) shows that at 1200°C, the sample is a single tetragonal phase. As the temperature increases to 1300°C, the gap between the two tetragonal peaks decreases and the two peaks eventually merge at 1400°C. However, the presence of the tetragonal structure is still noticeable in the peak. It is not until 1500°C that the structure becomes a single cubic fluorite phase. The volume expansion from the fluorite phase to body-centered tetragonal is approximately 0.7 %. The transformation was not seen to be destructive to the specimen on the macroscale.

The phase stability of the CCFOs is now compared to YSZ. The phase transition temperature (~1400°C) of 4L (stabilizer concentration = 5.8 at. %) is significantly lower than its 3YSZ counterpart (~ 2200°C). Also, the tetragonal phase is stable in 8YSZ from 600°C to 1300°C. Within the CCFOs, no phase change was detected down to 1200°C. Thus, the cubic phase stable region is enlarged in CCFOs in comparison with YSZ.

A proposed phase diagram of the $(Hf_{1/3}Zr_{1/3}Ce_{1/3})_{1-x}(Y_{1/2}Yb_{1/2})_xO_{2-\delta}$ CCFOs is shown in Figure 5. An approximated curved tetragonal-cubic boundary is drawn, which resembles that in the YSZ ($ZrO_2$-$Y_2O_3$) phase diagram [43]. Similar to YSZ, the increased stability of the cubic phase may be attributed to higher total fraction of cubic stabilizers that increases oxygen vacancy concentration, so that the tetragonal-cubic boundary is curved. Interestingly, no two-phase region has been observed between the tetragonal and cubic single-phase regions in the $(Hf_{1/3}Zr_{1/3}Ce_{1/3})_{1-x}(Y_{1/2}Yb_{1/2})_xO_{2-\delta}$ CCFOs, which differs from a wide two-phase region in the YSZ system. It is possible that such a two-phase region is too narrow (to be detected by our experiments) in this CCFO system.

In comparison with the YSZ phase diagram [43], it appears that the cubic fluorite phase is stabilized in $(Hf_{1/3}Zr_{1/3}Ce_{1/3})_{1-x}(Y_{1/2}Yb_{1/2})_xO_{2-\delta}$ CCFOs to a significantly greater degree by pushing the tetragonal envelope inward by at least a few hundred degrees at all temperatures. It is also known that Ce may enhance the stability of the cubic phase [54,55]. However, the $HfO_2$-$CeO_2$-$ZrO_2$ ternary phase diagram shows the three-phase region (monoclinic + tetragonal + cubic) for the equimolar composition at 1500°C [56]. It appears that the cubic stabilizers in this system affects the phase stability to a more considerable degree in this CCFO system than yttria does in the zirconia system. We should note that entropy stabilization of higher symmetry phases is seen in other reports [57,58]. This is consistent with the enhanced stability of the high-symmetry cubic phase in this CCFO system at lower stabilizer concentrations compared to YSZ.

## 4.4 Microstructure

Polished cross-sections were examined by SEM-EDS to investigate the microstructure and homogeneity on the microscale. All the compositions following the furnace anneal were solid compacts free of cracks with a random distribution of (low) porosity. All specimens are chemically homogeneous. SEM micrograph and the corresponding EDS maps on 4L, 4M, and 4H were selected as an example and shown in Figure 6. SEM-EDS analysis results of other compositions are shown in Figure S2 in Suppl. Data. All the CCFOs exhibited homogeneously distributed elements besides 5H (and 7L to a minor extent).

## 4.5 Mechanical Properties

The Young's modulus (*E*) and nanohardness data are shown in Table 1. Our measurements of our 8YSZ system agree well with literature values on both properties [59,60]. The composition series 5 $(Hf_{1/3}Zr_{1/3}Ce_{1/3})_{1-x}(Y_{1/2}Ca_{1/2})_xO_{2-\delta}$ specimens had significantly lower modulus while the rest of the CCFOs had similar values. Calcium is a large ion (ionic radius = 1.12 Å) with a different charge (2+) than the rest of the cations investigated (3+/4+), which may result in the different modulus.

Nanoindentation revealed similar nanohardness values of all compositions to be 16-20 GPa, on par with that of 8YSZ (17.3 ± 1.5 GPa). Specimen 5M $(Hf_{0.284}Zr_{0.284}Ce_{0.284}Y_{0.074}Ca_{0.074})_xO_{2-\delta}$ is the hardest (19.6 ± 0.6 GPa), despite its lowest modulus. A representative load-displacement curve for each specimen is shown in Figure S3 in Suppl. Data.

## 4.6 Thermal Conductivity

One of the most attractive properties of YSZ is its low thermal conductivity (*k*). Thermal conductivity measurements were performed from room temperature to 1000°C and are shown in Figure 7, with the raw diffusivity and heat capacity data shown in Figures S4 and S5 in Suppl. Data. Noting that the measured room-temperature *k* values for 4H and 7H are greater than those previously reported [9]. The differences may arise from differences in measurement methods (laser flash analysis vs. time-domain thermoreflectance). The measured temperature-dependent thermal conductivity of 8YSZ is in an excellent agreement with literature [61]. For some specimens (e.g., 8YSZ, 4H and 7H), thermal conductivity initially decreases but then slightly increases with temperature above around 800°C. This increasing trend is presumably due to radiation heat transfer, as also observed and speculated in the literature [62–64]. If the radiation thermal conductivity is significant, the data reduction method used in the laser flash analysis is no longer fully valid. Therefore, the data shown in the high-temperature range (e.g., above ~800°C) should be taken as an approximation only. Here, the electronic conduction to the thermal conductivity can be ignored, as YSZ and similar oxides are largely insulating [65,66].

The temperature-dependent thermal conductivity values give insight into the thermal conduction mechanism. 8YSZ exhibits a constant $k$ around 2 W $m^{-1}$ $K^{-1}$ but increases slightly at $T \geq 800$°C due to radiation heat transfer. From room temperature to 1000°C, the reduction in $k$ for the CCFOs compared to 8YSZ is about 20 %. This reduction of the thermal conductivity is likely originated from the increased phonon scattering. Both equimolar 4H and 7H show the classical $1/T$ relationship representative of Umklapp scattering at low temperatures, which then rise at higher temperatures from radiative effects (like 8YSZ).

Notably, at lower stabilizer concentrations, the temperature dependency changes to an amorphous-like or disordered trend [67], where the thermal conductivity increases slowly with temperature or nearly temperature-independent. Disordered trends arise when the propagating phonon modes are severely suppressed, and the main heat conduction mode becomes non-propagating diffusons [68,69]. It is interesting to note that the equimolar compositions, which have the most disorder on the cation sublattice, do not display this trend. Thus, this amorphous-like behavior is probably a result of oxygen vacancies. Oxygen vacancies are known to cluster in the fluorite structure to form bi- and tri-vacancies when the oxygen vacancy concentration increases above 5 % [51,53,70,71]. The clustering of the vacancies may reduce phonon scattering of the system [72]. We should note that the effects of oxygen vacancies are coupled with mass disorder and lattice distortion (local strains) to influence thermal conductivity. Specifically, the presence of disordered vacancies may increase the distortion on the oxygen sublattice according to the point defect scattering theory [73–75]. Consistently, equimolar 4H and 7H display a crystalline-like conductivity that has a $1/T$ dependency with temperature, while other compositions exhibit amorphous-like behaviors.

This also explains an important finding of this study that the non-equimolar medium-entropy CCFOs exhibit even lower (and temperature-independent) thermal conductivities than their equimolar high-entropy counterparts. This is clearly demonstrated, e.g., in Figure 8 for the Composition 4 series $(Hf_{1/3}Zr_{1/3}Ce_{1/3})_{1-x}(Y_{1/2}Yb_{1/2})_xO_{2-\delta}$ as an example.

The thermal conductivities of 5L and 5M $(Hf_{1/3}Zr_{1/3}Ce_{1/3})_{1-x}(Y_{1/2}Ca_{1/2})_xO_{2-\delta}$ specimens are lower than those of the composition 7 series $(Hf_{1/3}Zr_{1/3}Ce_{1/3})_{1-x}(Y_{1/2}Gd_{1/2})_xO_{2-\delta}$ specimens, which are slightly lower than that of the Composition 4 series $(Hf_{1/3}Zr_{1/3}Ce_{1/3})_{1-x}(Y_{1/2}Yb_{1/2})_xO_{2-\delta}$ specimens. The introduction of $Ca^{2+}$ reduced the conductivity of the system further. As shown in Figure 7, Composition 5M approaches the minimum thermal conductivity limit (e.g., Cahill limit [76]) at high temperature but there is still room for further reduction. A factor may be the size disorder. Composition series 5, 7, and 4 contain $Ca^{2+}$, $Gd^{3+}$, and $Yb^{3+}$, which have Shannon effective ionic radii of 1.12 Å, 1.053 Å, and 0.985 Å, respectively [77]. This suggests size disorder as a possible factor in controlling thermal conductivity. In this case, the different charge state of $Ca^{2+}$ plays an indirect effect.

### 4.7 Stiffness to Conductivity (*E/k*) Ratio

Interestingly, the mechanical properties of the CCFOs are reasonably similar to YSZ, but the thermal conductivity is reduced. Traditionally, the modulus and conductivity are directly related because strong bonding increases both [21]. The room temperature *k*, *E*, and *E/k* for the Composition 4 series $(Hf_{1/3}Zr_{1/3}Ce_{1/3})_{1-x}(Y_{1/2}Yb_{1/2})_xO_{2-\delta}$ are shown in Figure 8. Here, there is a clear improvement in the *E/k* ratio when we move from equimolar high-entropy 4H to non-equimolar concentrations. A sharp drop in modulus in 4L and consequently *E/k* may be due to the relatively weaker tetragonal structure compared to cubic. Thus, the highest *E/k* ratio of 133.6 ± 4.3 $GPa \cdot m \cdot K \cdot W^{-1}$ is achieved for 4M $Hf_{0.284}Zr_{0.284}Ce_{0.284}Y_{0.074}Yb_{0.074})_xO_{2-\delta}$.

The properties of all CCFOs are summarized in Table 1. The same trend stated above is seen in the other composition series as well, where the non-equimolar CCFOs (variant "M") are generally more thermally-insulative (with lower thermal conductivity) compared to their equimolar high-entropy counterparts (variant "H").

Generally, as the concentration of oxygen vacancy concentration increases (from "L" to "H"), so does the thermal conductivity and modulus to an extent. The *k* appears to reach a minimum (and *E/k* reaches a maximum) around the composition of variant "M" which has a nominal oxygen vacancy percentage of 3.7 %. However, this trend is not upheld in Composition 5 series $(Hf_{1/3}Zr_{1/3}Ce_{1/3})_{1-x}(Y_{1/2}Ca_{1/2})_xO_{2-\delta}$ specimens because $Ca^{2+}$ introduces more oxygen vacancies in comparison with $Yb^{3+}$ and $Gd^{3+}$. Note that 5L $Hf_{0.314}Zr_{0.314}Ce_{0.314}Y_{0.029}Ca_{0.029})_xO_{2-\delta}$ has a nominal oxygen vacancy concentration of 2.2 %, while 5M $Hf_{0.284}Zr_{0.284}Ce_{0.284}Y_{0.074}Ca_{0.074})_xO_{2-\delta}$ has 5.6 %. Overall, an oxygen vacancy concentration around 3 % should display the most improved properties in terms of the *E*/*k* ratio. This shows the importance of oxygen vacancies to the thermal properties of this new class of CCFOs.

Notably, most CCFOs display a 5 – 30 % increase in *E/k* ratio compared to YSZ. This study further showed that one can explore medium non-equimolar CCFOs to reduce *k* (beyond their equimolar high-entropy counterparts) while retaining the modulus (at least to certain concentration) and hardness to achieve higher *E*/*k* ratio. This allows for a new direction to design stiff and insulative ceramics. It is important to note not only HECs represent a new compositional space, but also most medium-entropy and/or non-equimolar compositions are largely unexplored by any prior studies (i.e., in neither the traditional ceramic research nor in the most recent HEC studies); yet, they represent a vast unexplored compositional space.

## 5 Conclusions

This study broadens high-entropy ceramics (HECs) to compositionally-complex ceramics (CCCs), which can be alternatively and equivalently called "multi-principal cation ceramics (MPCCs)," to explore medium-entropy non-equimolar compositions using compositionally-complex fluorite oxides (CCFOs) as an exemplar. Nine compositions CCFOs with the general formula of $(Hf_{1/3}Zr_{1/3}Ce_{1/3})_{1-x}(Y_{1/2}X_{1/2})_xO_{2-\delta}$ (where $X$ = Yb, Ca, and Gd) have been fabricated and eight of exhibit single solid-solution phases at 1500 °C.

In comparison with YSZ, these CCFOs exhibit increased cubic stability and lower thermal conductivity, while retaining comparable modulus and hardness. Moreover, the temperature-dependent thermal conductivity in the non-equimolar CCFOs shows an amorphous-like behavior.

Most notably, the medium-entropy non-equimolar CCFOs exhibit even lower thermal conductivity (*k*) and higher *E*/*k* ratios than their equimolar high-entropy counterparts. Thus, these results suggest a new direction to design, fabricate, and test insulative yet stiff CCFOs specifically, as well as other high-performance CCCs or MPCCs in general, via exploring non-equimolar and/or medium-entropy compositions beyond HECs.

This study has also demonstrated the importance of oxygen vacancies, particularly their clustering, in influencing the thermal conductivity, which is the underlying reason that non-equimolar medium-entropy CCFOs outperform their equimolar high-entropy counterparts.

**Acknowledgment:** This material is primarily based upon work supported by the U.S. Department of Energy's Office of Energy Efficiency and Renewable Energy (EERE) under Solar

Energy Technologies Office (SETO) Agreement Number EE0008529 (for Feb 2019 to July 2020). J.L. and A.W. also acknowledge earlier (before Feb 2019) and partial support of a Vannevar Bush Faculty Fellowship (via ONR Grant No. N00014-16-1-2569) and an associated Laboratory-University Collaboration Initiative (LUCI) program to investigate interfaces in thermal and environmental barrier coatings, which subsequently led to this new direction of HEFOs and CCFOs. We would like to thank Dr. D. Cole at the Army Research Laboratory (ARL) for providing the facility and assistance for the nanoindentation experiments (that led to the nanohardness data in Table 1), and A.W. also acknowledges an ARL summer internship that allowed him to use the nanoindentation facility at ARL.

**Appendix A. Supplementary Data**

Supplementary material related to this article can be found, in the online version, at doi: xxxxxxxx

**Table 1.** Summary of properties measured at room temperature for all single-phase specimens annealed at 1500°C for 24 hours. Specimen 3YSZ and 5H have multiple phases, so their properties were not measured. Nanohardness was not measured for 4H due to an experimental issue.

| Identifier | Composition | $a_c$ (Å) | $a_t$ (Å) | $c_t$ (Å) | Microstrain (%) | Theoretical Density (g/cm$^3$) | Relative Density (%) | Nanohardness ($GPa$) | $E$ ($GPa$) | $k$ $\left(\frac{W}{m \cdot K}\right)$ | $E/k$ $\left(\frac{GPa \cdot m \cdot K}{W}\right)$ |
|---|---|---|---|---|---|---|---|---|---|---|---|
| 3YSZ | $Zr_{0.942}Y_{0.058}O_{2-\delta}$ | — | — | — | — | — | — | — | — | — | — |
| 8YSZ | $Zr_{0.852}Y_{0.148}O_{2-\delta}$ | 5.134 | — | — | 0.15 | 5.97 | 97.3 ± 0.6 | 17.3 ± 1.5 | 219.2 ± 2.6 | 2.19 ± 0.07 | 100.3 ± 3.2 |
| 4L | $Hf_{0.314}Zr_{0.314}Ce_{0.314}Y_{0.029}Yb_{0.029}O_{2-\delta}$ | — | 3.682 | 5.238 | 0.19 | 7.85 | 96.9 ± 0.6 | 17.1 ± 1.8 | 201.2 ± 2.2 | 1.78 ± 0.05 | 113.0 ± 3.6 |
| 4M | $Hf_{0.284}Zr_{0.284}Ce_{0.284}Y_{0.074}Yb_{0.074}O_{2-\delta}$ | 5.210 | — | — | 0.16 | 7.83 | 97.5 ± 0.6 | 17.7 ± 1.7 | 232.8 ± 2.2 | 1.74 ± 0.05 | 133.6 ± 4.3 |
| 4H | $Hf_{0.2}Zr_{0.2}Ce_{0.2}Y_{0.2}Yb_{0.2}O_{2-\delta}$ | 5.227 | — | — | 0.08 | 7.59 | 98.2 ± 0.6 | — | 232.6 ± 2.4 | 2.23 ± 0.07 | 104.4 ± 3.3 |
| 5L | $Hf_{0.314}Zr_{0.314}Ce_{0.314}Y_{0.029}Ca_{0.029}O_{2-\delta}$ | 5.214 | — | — | 0.16 | 7.67 | 97.3 ± 0.6 | 18.4 ± 1.3 | 180.1 ± 4.8 | 1.65 ± 0.05 | 109.4 ± 4.4 |
| 5M | $Hf_{0.284}Zr_{0.284}Ce_{0.284}Y_{0.074}Ca_{0.074}O_{2-\delta}$ | 5.218 | — | — | 0.10 | 7.30 | 95.4 ± 0.6 | 19.6 ± 0.6 | 153.8 ± 2.3 | 1.54 ± 0.05 | 99.6 ± 3.4 |
| 5H | $Hf_{0.2}Zr_{0.2}Ce_{0.2}Y_{0.2}Ca_{0.2}O_{2-\delta}$ | — | — | — | — | — | — | — | — | — | — |
| 7L | $Hf_{0.314}Zr_{0.314}Ce_{0.314}Y_{0.029}Gd_{0.029}O_{2-\delta}$ | — | 3.690 | 5.245 | 0.17 | 7.78 | 96.1 ± 0.6 | 18.5 ± 1.4 | 199.4 ± 2.0 | 1.74 ± 0.05 | 114.3 ± 3.6 |
| 7M | $Hf_{0.284}Zr_{0.284}Ce_{0.284}Y_{0.074}Gd_{0.074}O_{2-\delta}$ | 5.225 | — | — | 0.08 | 7.70 | 96.9 ± 0.6 | 18.2 ± 1.7 | 218.9 ± 2.3 | 1.70 ± 0.05 | 128.4 ± 4.2 |
| 7H | $Hf_{0.2}Zr_{0.2}Ce_{0.2}Y_{0.2}Gd_{0.2}O_{2-\delta}$ | 5.265 | — | — | 0.07 | 7.28 | 98.2 ± 0.6 | 16.3 ± 1.2 | 238.9 ± 2.7 | 2.19 ± 0.06 | 106.3 ± 3.3 |

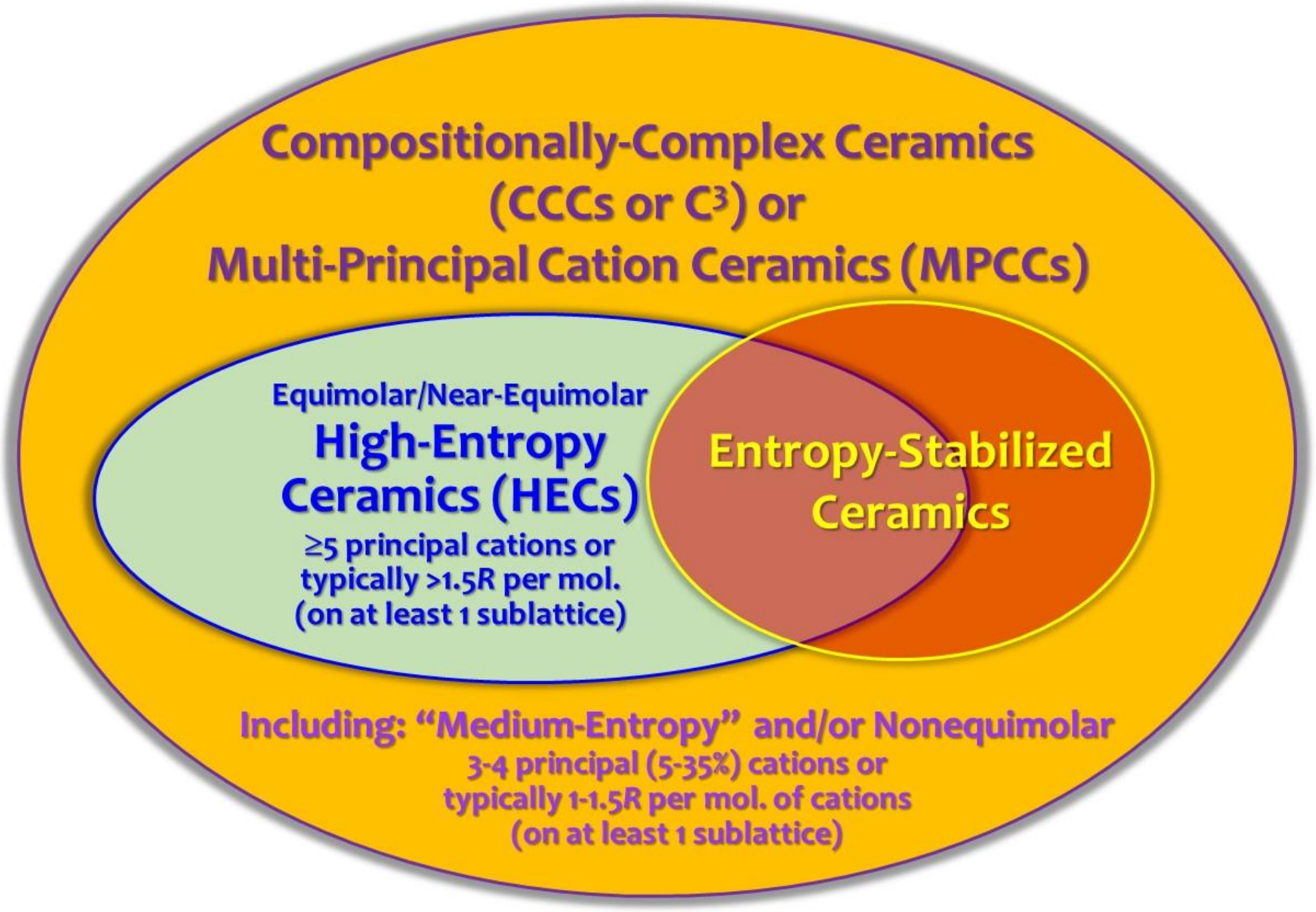

**Figure 1.** Schematic of proposed compositionally-complex ceramics (CCCs or $C^3$), which can alternatively and equivalently be named as "multi-principal cation ceramics (MPCCs)." CCCs include high-entropy ceramics (HECs) and entropy-stabilized ceramics as subsets; moreover, CCCs also include medium-entropy and/or non-equimolar compositions that are not qualified as HECs. Following the definitions of metallic high-entropy alloys (HEAs), HECs typically refer to equimolar or near-equimolar compositions of five or more principal cations (usually producing >1.5 *R* per mol. of cations ideal configuration entropy) on at least one sublattice. We propose to broaden HECs to CCCs to further include non-equimolar and/or medium-entropy compositions of three or four principal (typically 5%-35%) cations or generally 1-1.5 *R* per mol. ideal configuration entropy on at least one sublattice. Note that different (mostly subjective) criteria exist so that the boundaries of high- or medium-entropy compositions are only loosely defined. We should further note that most non-equimolar and/or medium-entropy ceramics (based on the above definitions) also represent a largely unexplored compositional space (in addition to HECs).

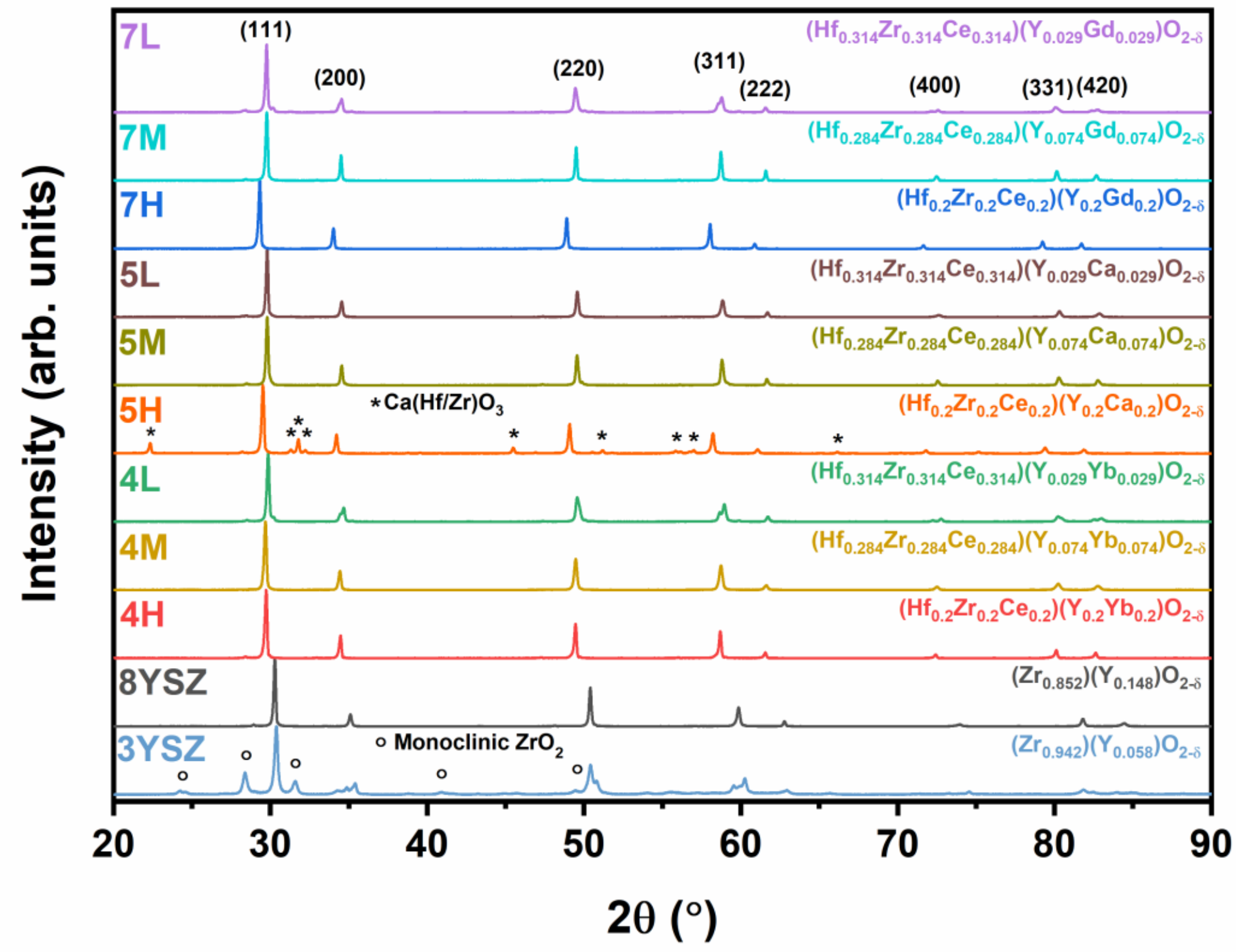

**Figure 2.** XRD patterns of all 11 specimens (nine CCFOs, along with 8YSZ and 3YSZ) annealed at 1500°C for 24 hours and furnace cooled. All variants "M" and "H" specimens, except 5H, exhibit single solid-solution phases of the fluorite structure. In variant "L" specimens, both 4L and 7L exhibit the tetragonal structure, while 5L retains the cubic fluorite structure.

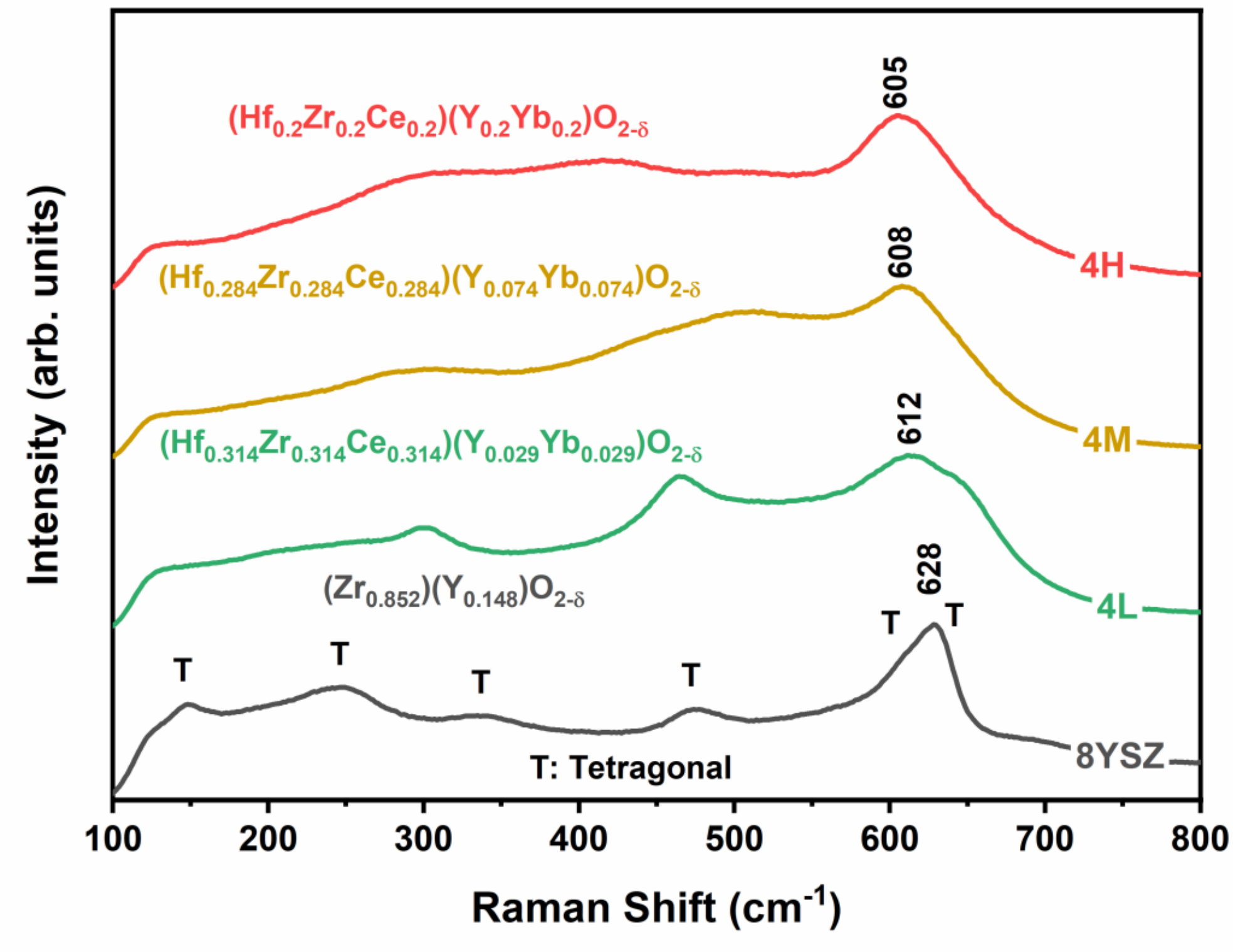

**Figure 3.** Raman spectra showing some degrees of tetragonality present in selected specimens annealed at 1500°C (albeit that some are shown as cubic fluorite by XRD). As the stabilizer concentration is increased, the degree of the tetragonality decreases.

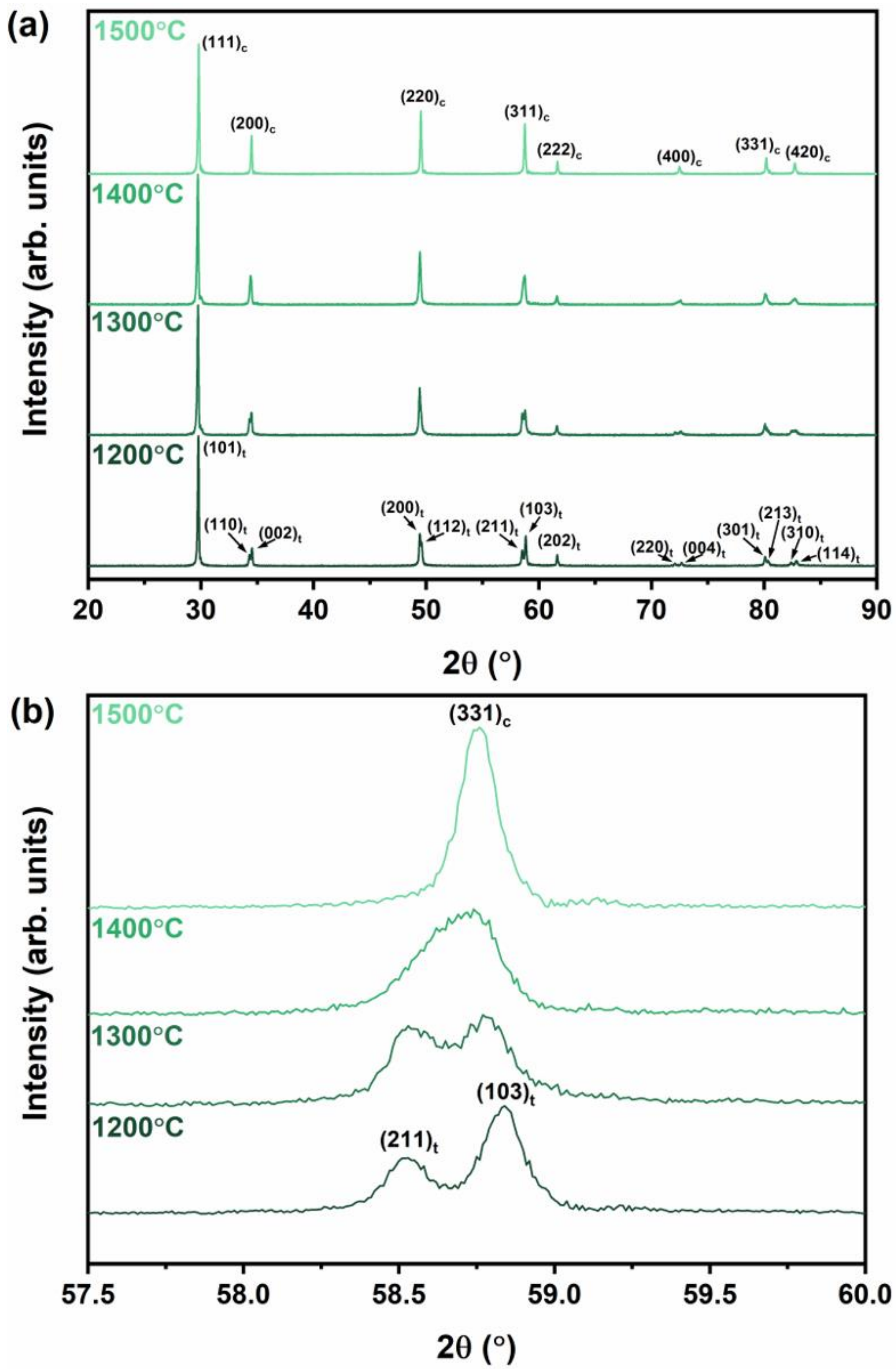

**Figure 4.** (a) XRD spectra of four 4L ($Hf_{0.314}Zr_{0.314}Ce_{0.314}Y_{0.029}Yb_{0.029}O_{2-\delta}$) specimens produced by 24-hour annealing at different temperatures (labeled in the figure) and subsequent quenching. A phase transformation occurs around 1400°C and completes at 1500°C. (b) Expanded views of the XRD spectra showing the transition from two tetragonal peaks to a single cubic peak.

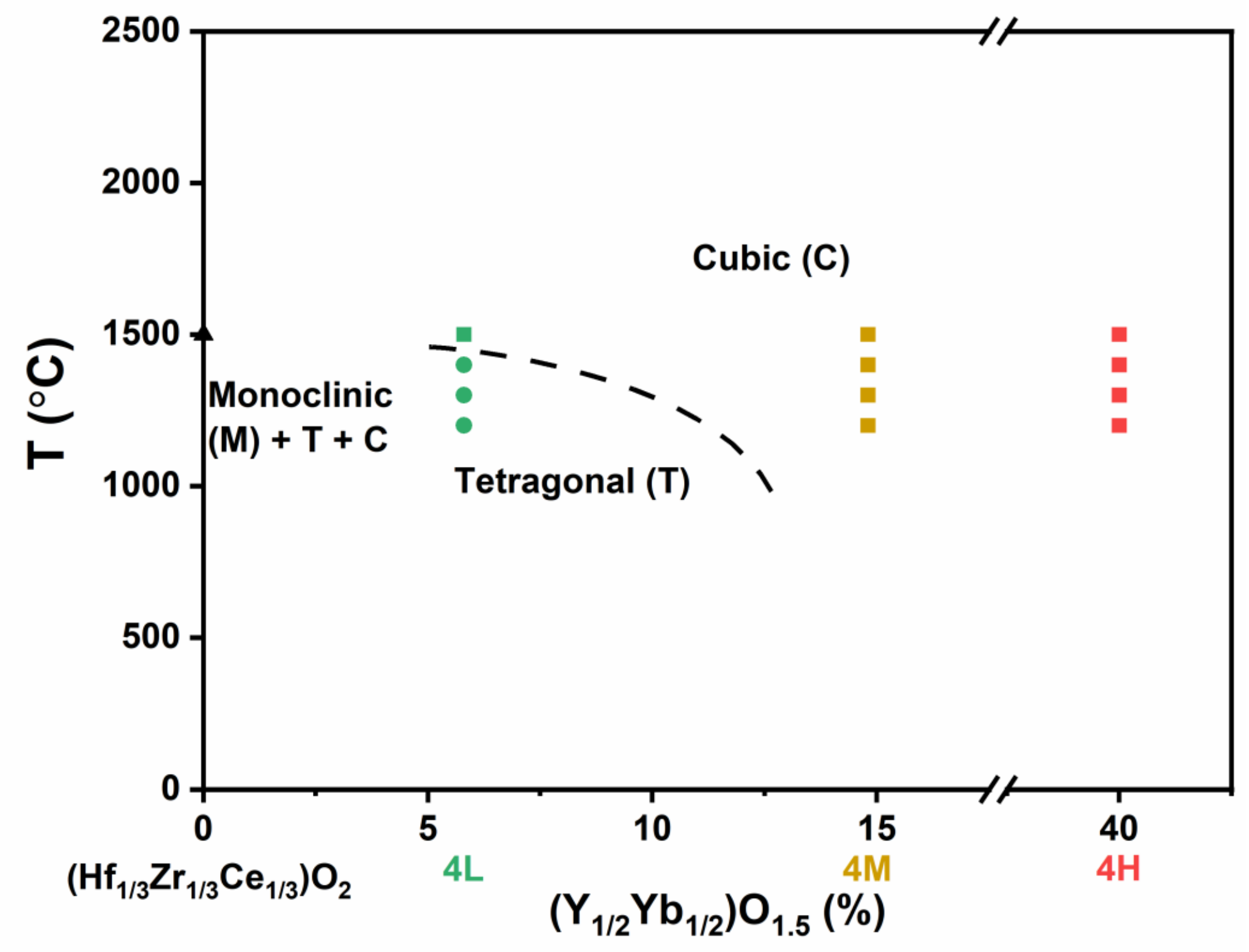

**Figure 5.** Proposed phase diagram of specimens $(Hf_{1/3}Zr_{1/3}Ce_{1/3})_{1-x}(Y_{1/2}Yb_{1/2})_xO_{2-\delta}$ (4L, 4M, and 4H). Composition 4L undergoes a transformation from the tetragonal to cubic phase at ~1400°C – 1500°C while 4M and 4H remained in the cubic fluorite phase from 1200°C to 1500°C. All specimens were annealed at the respective temperatures for 24 h and subsequently quenched. The composition $(Hf_{1/3}Zr_{1/3}Ce_{1/3})O_2$ exhibits three phases according to Andrievskaya *et al.*[56]. (Legends: ▲ = 3 phase; ● = tetragonal; ■ = cubic.)

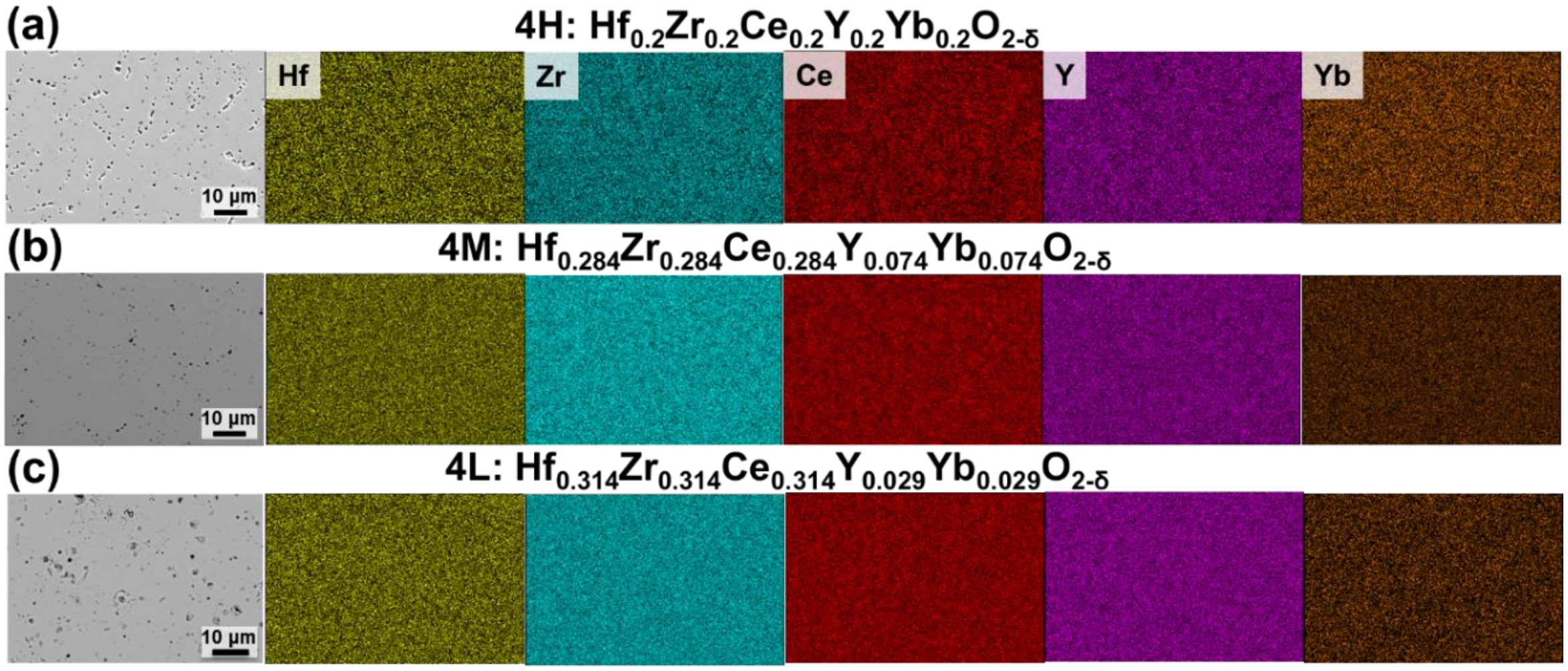

**Figure 6.** SEM images and corresponding EDS elemental maps on the cross sections of 4H, 4M, and 4L annealed at 1500°C for 24 hours. The cation distributions are uniform in all specimens at the microscale.

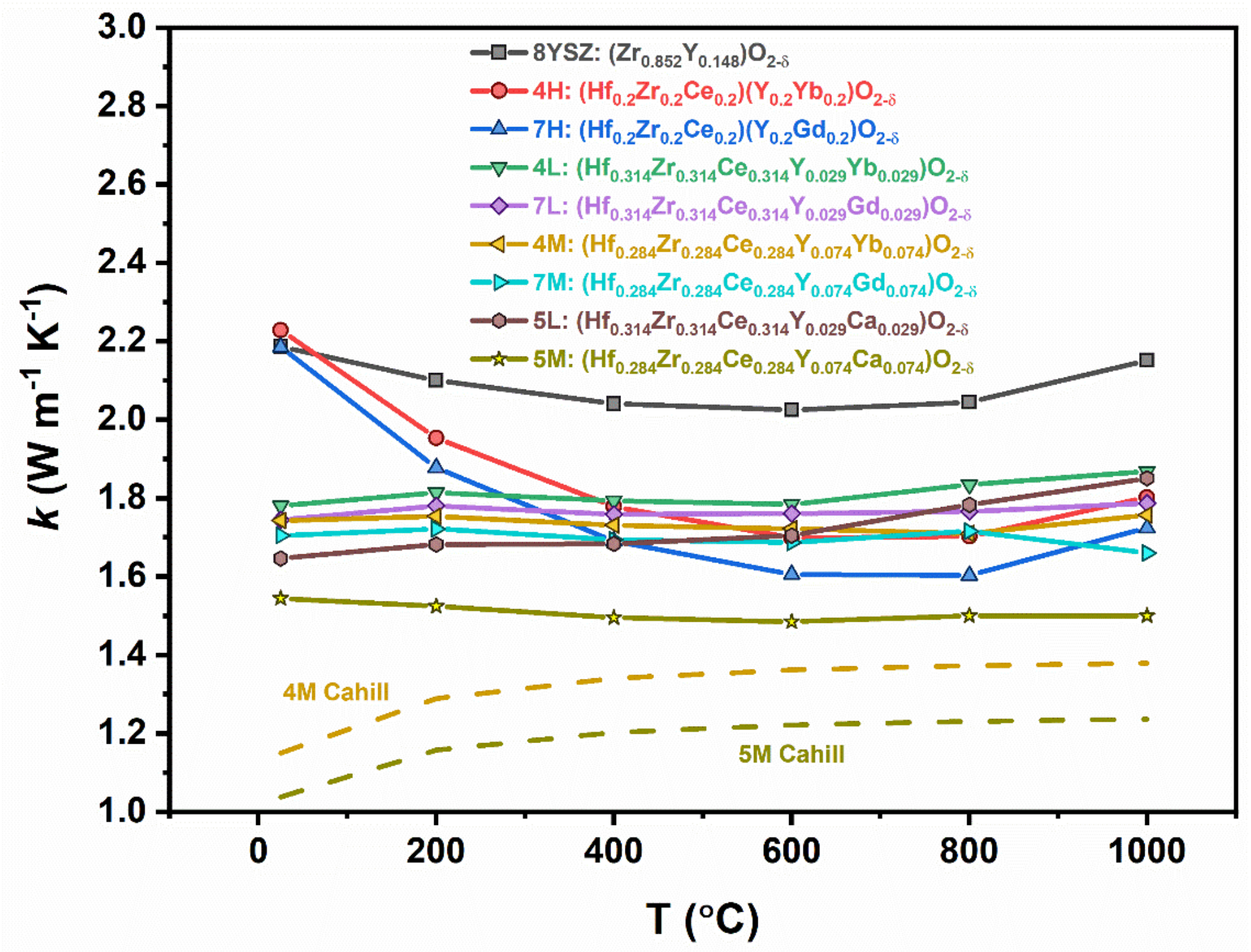

**Figure 7.** Measured thermal conductivity vs. temperature curves for all single-phase specimens from room temperature to 1000°C. The calculated phonon limits for 4M and 5M by using the Cahill [76] model are shown by the dashed lines. All specimens were annealed at 1500°C for 24 hours prior to the thermal conductivity measurements.

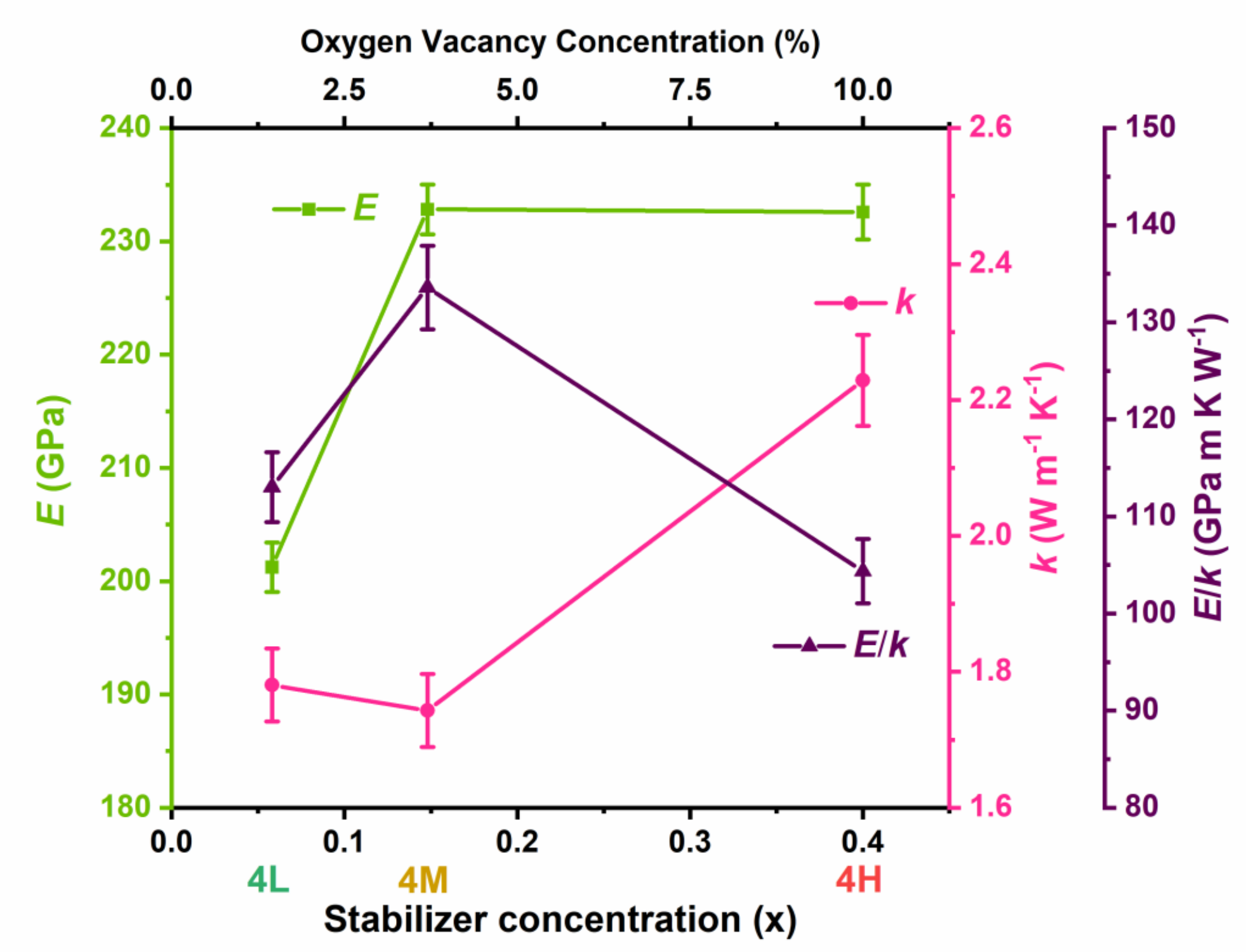

**Figure 8.** Room-temperature Young's modulus (*E*), thermal conductivity (*k*), and *E/k* ratio for $(Hf_{1/3}Zr_{1/3}Ce_{1/3})_{1-x}(Y_{1/2}Yb_{1/2})_xO_{2-\delta}$ series of specimens 4L, 4M, and 4H. 4M exhibits the lowest thermal conductivity and the highest $E/k$ ratio. All specimens were annealed at 1500°C for 24 hours.

# Appendix A:

# Supplementary Data

## From High-Entropy Ceramics to Compositionally-Complex Ceramics: A Case Study of Fluorite Oxides

Andrew J. Wright[a], Qingyang Wang[b], Chuying Huang[a], Andy Nieto[c], Renkun Chen[b], Jian Luo[a,*]

[a]Department of NanoEngineering, University of California, San Diego, La Jolla, California, USA

[b]Department of Mechanical & Aerospace Engineering, University of California, San Diego, La Jolla, California, USA

[c]Department of Mechanical and Aerospace Engineering, Naval Postgraduate School, Monterey, California, USA

*E-mail address: jluo@alum.mit.edu

Figure S1 shows the XRD patterns of 4L, 4M and 4H air-quenched from 1200°C to 1500°C, along with those furnace-cooled from 1500°C. Composition 4L has a BCT to FCC phase transformation around 1400°C. Both 4M and 4H remained cubic for all examined quenching temperatures.

The microstructure and corresponding elemental distribution maps for 5H, 5M, 5L, 7H, 7M, and 7L specimens (furnace-cooled from 1500°C) are shown in Figure S2. There is a secondary $Ca(Hf,Zr)O_3$ phase in 5H and slight Ce deficiencies in 7L. All other specimens show homogenously distributed elements.

A representative load-displacement curve from nanoindentation for each specimen furnace-cooled from 1500°C is displayed in Figure S3.

The measured thermal diffusivities and calculated heat capacities from room temperature up to 1000°C for the eight single-phase specimens (including 8YSZ) are shown in Figures S3 and S4, which were used to calculate thermal conductivity *vs.* temperature curves in Figure 7 in the main text.

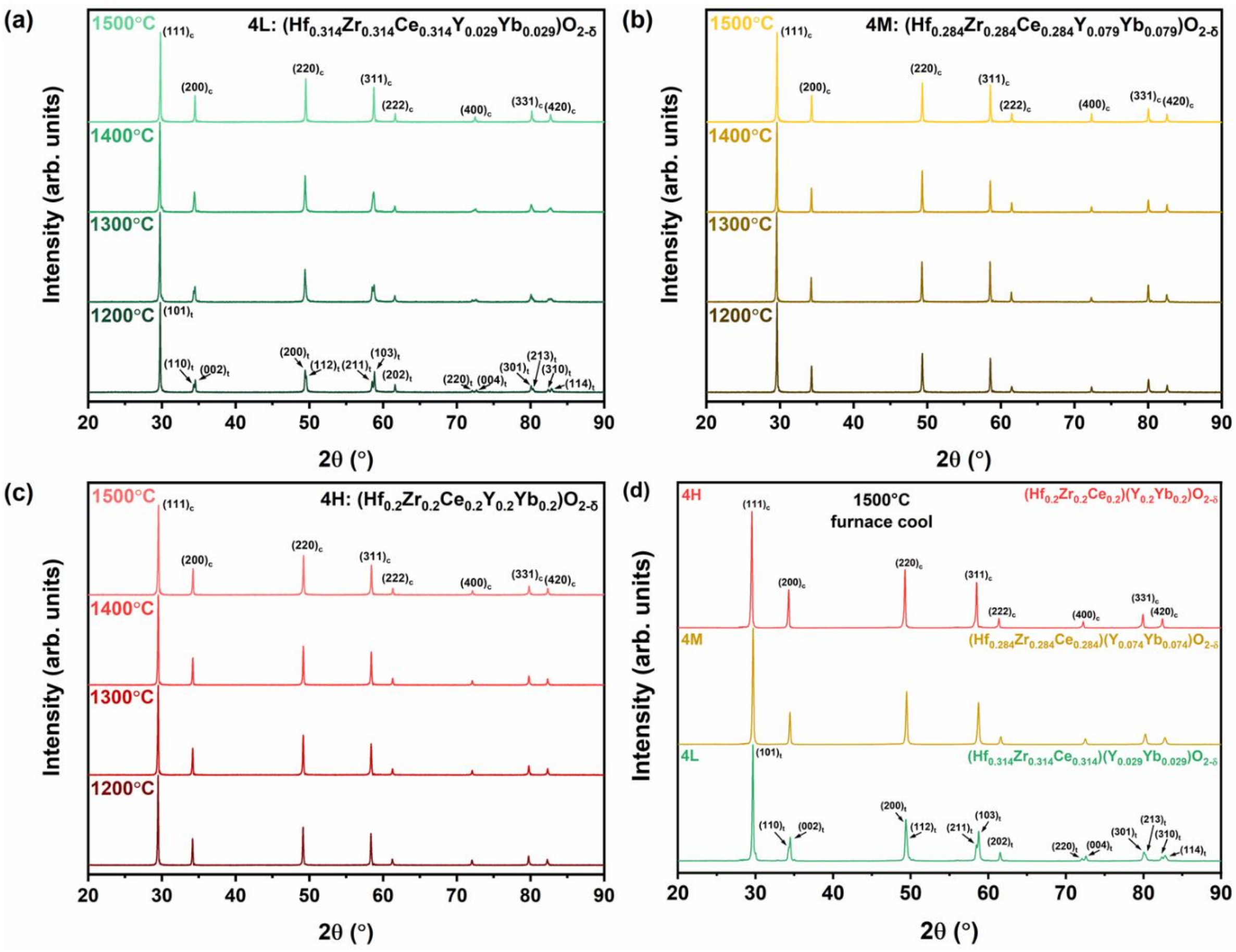

**Figure S1.** XRD patterns of the specimens air-quenched from 1200°C - 1500°C for compositions **(a)** 4L, **(b)** 4M, and *(c)* 4H. **(d)** The XRD patterns of three specimens of 4L, 4M, and 4H furnace-cooled from 1500°C as reference.

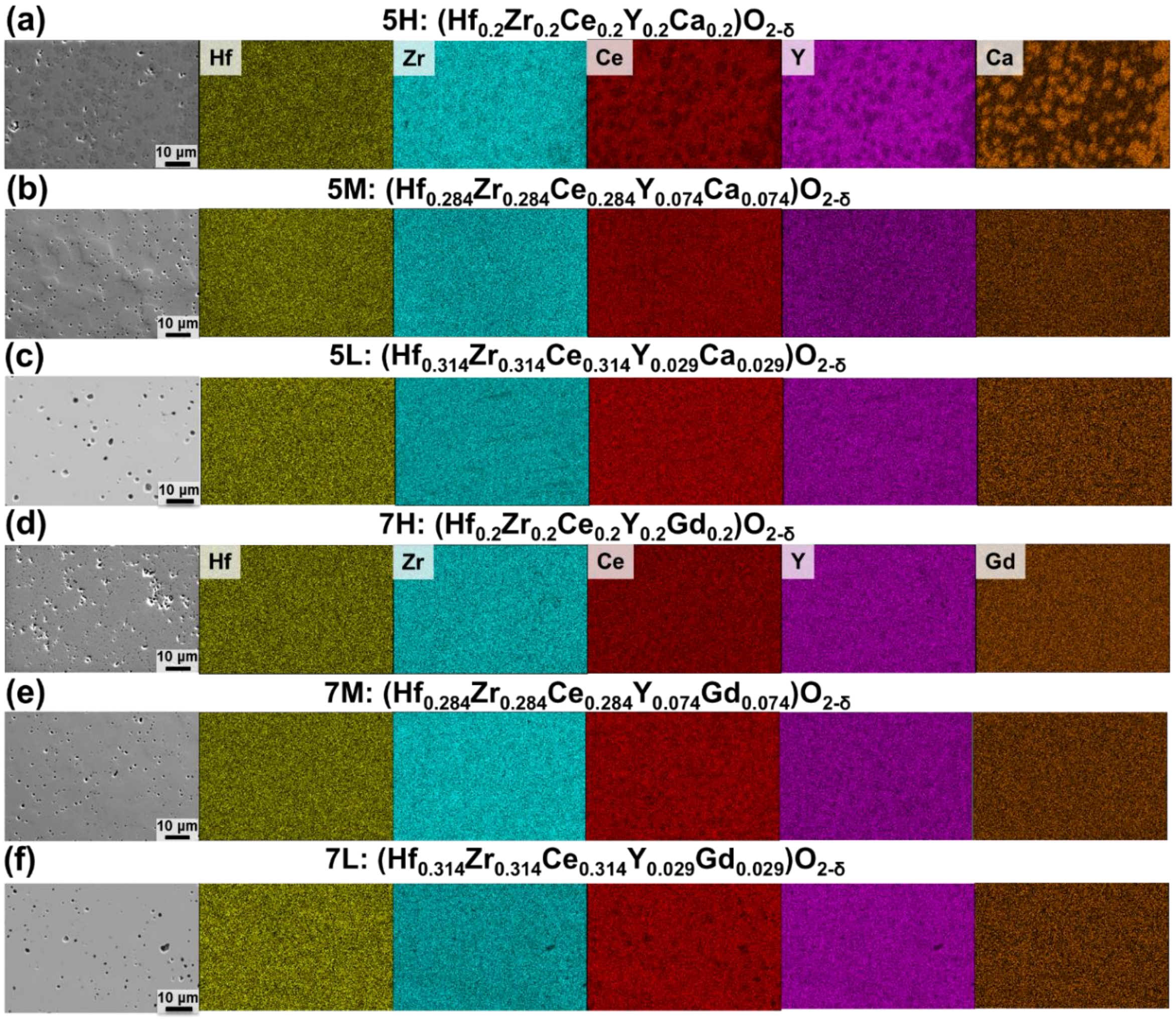

**Figure S2.** SEM images and EDS elemental maps of specimens **(a)** 5H, **(b)** 5M, **(c)** 5L, **(d)** 7H, **(e)** 7M, and **(f)** 7L (furnace-cooled from 1500°C).

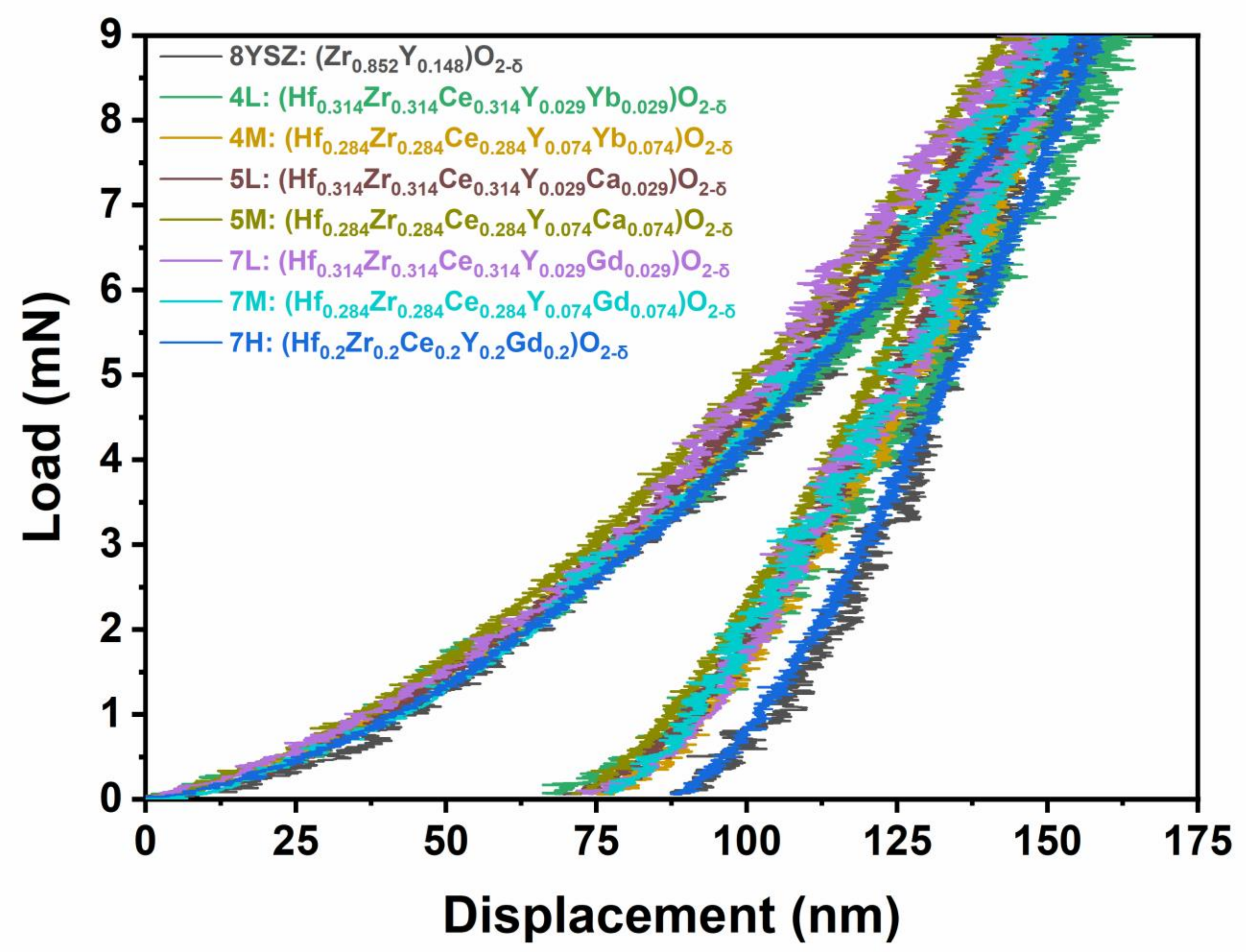

**Figure S3.** Characteristic load-displacement curves for each specimen (furnace-cooled from 1500°C).

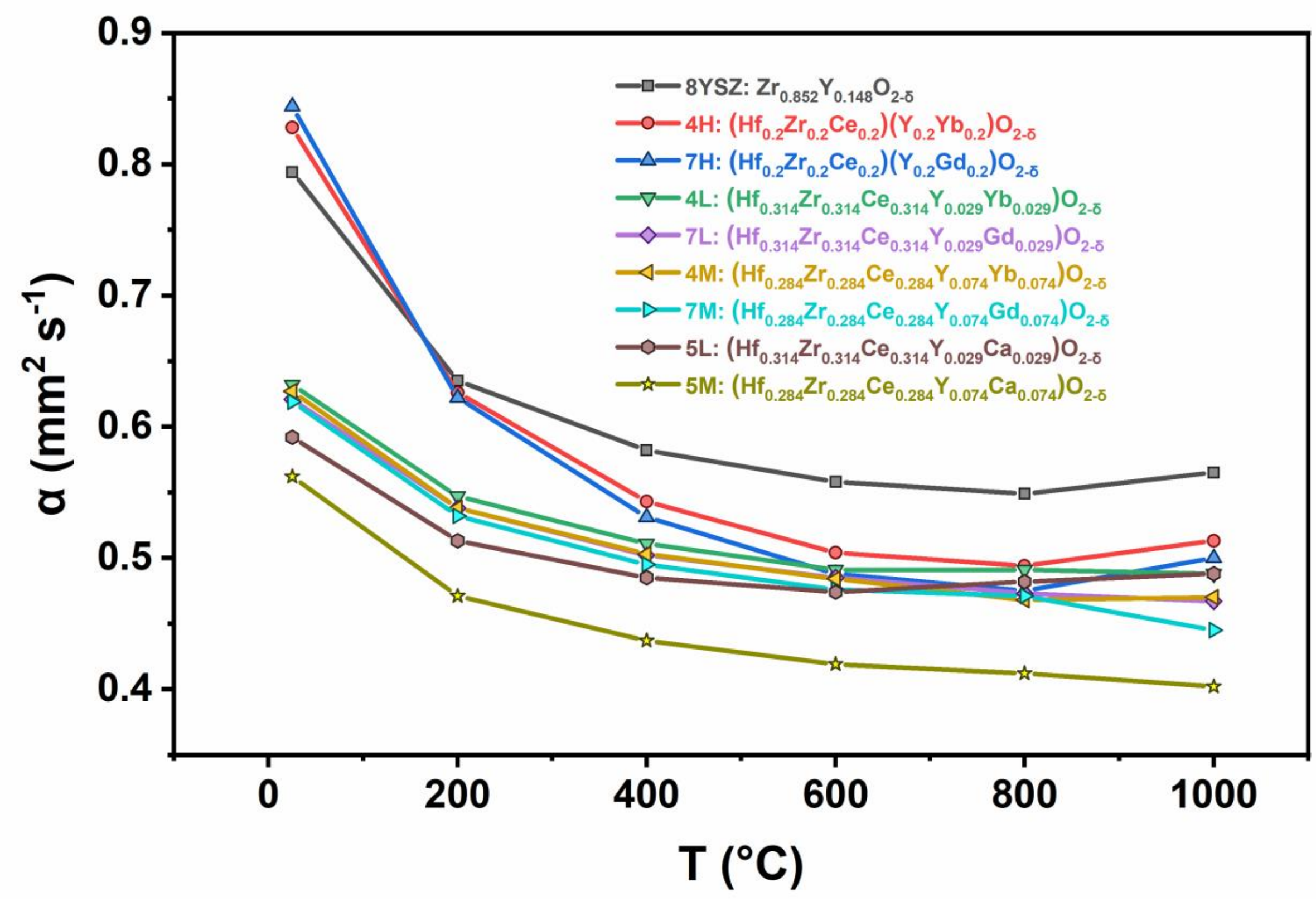

**Figure S4.** Measured thermal diffusivity *vs.* temperatures curves for all furnace-cooled single-phase specimens from room temperature to 1000°C.

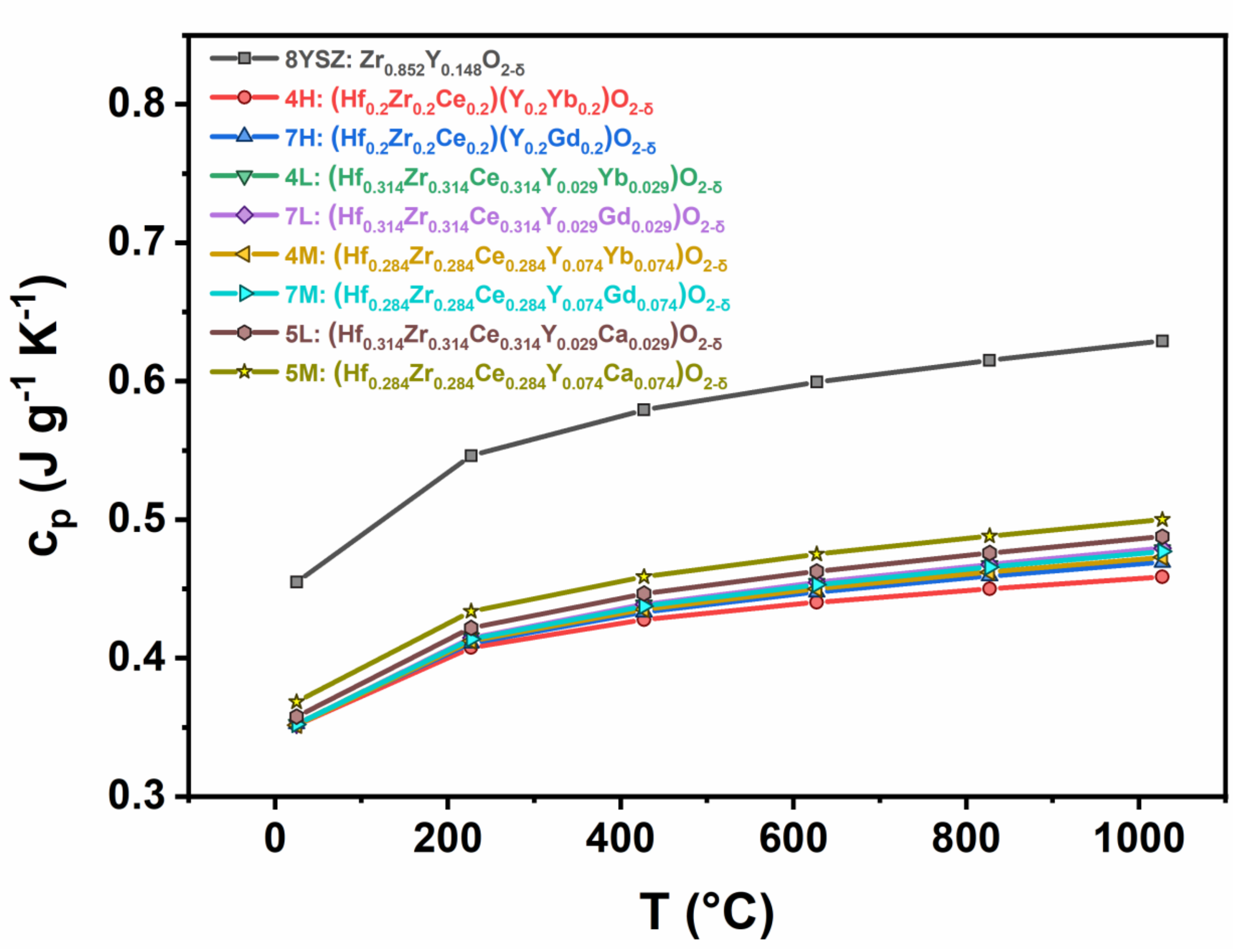

**Figure S5.** Calculated temperature-dependent heat capacity based on Neumann-Kopp law using the heat capacities of the constituent oxides from I. Barin, *Thermochemical Data of Pure Substances*, VCH, Weinheim, 1995.